\documentclass[aps,prl,english,twocolumn,letterpaper,notitlepage]{revtex4-2}
\pdfoutput=1
\usepackage[T1]{fontenc}
\usepackage{graphicx}
\usepackage{epstopdf}
\usepackage{amssymb}
\usepackage{amsmath}
\PassOptionsToPackage{hyphens}{url}
\usepackage[urlcolor=blue,hyperindex, breaklinks,colorlinks,bookmarks=true,linkcolor=red,citecolor=black]{hyperref}
\usepackage[normalem]{ulem}
\usepackage{color}
\usepackage{rotating}
\usepackage{physics}
\usepackage{bm}
\usepackage{soul}
\PassOptionsToPackage{linktocpage}{hyperref}

\usepackage[english]{babel}

\begin{document}

\title{Neuromorphic computing with a single qudit}

\author{W.~D.~Kalfus}
\altaffiliation{Current address: Department of Applied Physics, Yale University, New Haven, CT 06520, USA and Yale Quantum Institute, Yale University, New Haven, CT 06520, USA}
\email{william.kalfus@yale.edu}
\author{G.~J.~Ribeill}
\author{G.~E.~Rowlands}
\author{H.~K.~Krovi}
\author{T.~A.~Ohki}
\author{L.~C.~G.~Govia}
\email{luke.c.govia@raytheon.com}
\affiliation{Quantum Engineering and Computing, Raytheon BBN Technologies, 10 Moulton St., Cambridge, MA 02138, USA}

\begin{abstract}
  Accelerating computational tasks with quantum resources is a widely-pursued goal that is presently limited by the challenges associated with high-fidelity control of many-body quantum systems. The paradigm of reservoir computing presents an attractive alternative, especially in the noisy intermediate-scale quantum era, since control over the internal system state and knowledge of its dynamics are not required. Instead, complex, unsupervised internal trajectories through a large state space are leveraged as a computational resource. Quantum systems offer a unique venue for reservoir computing, given the presence of interactions unavailable in analogous classical systems, and the potential for a computational space that grows exponentially with physical system size. Here, we consider a reservoir comprised of a single qudit ($d$-dimensional quantum system). We demonstrate a robust performance advantage compared to an analogous classical system accompanied by a clear improvement with Hilbert space dimension for two benchmark tasks: signal processing and short-term memory capacity. Qudit reservoirs are directly realized by current-era quantum hardware, offering immediate practical implementation, and a promising outlook for increased performance in larger systems.
\end{abstract}

\maketitle

Harnessing the complex dynamics of quantum systems to accelerate information processing promises transformative improvements in performance, but in the standard methodology \cite{Nielsen10} requires robust control and resilience to noise and error. Ideas inspired by neuromorphic computing \cite{Schuman17} offer a possible path to circumventing these requirements, and are the focus of the growing domains of quantum neuromorphic computing \cite{Markovic:2020aa} and quantum machine learning \cite{Biamonte:2017aa,Dunjko:2018aa}. These are especially attractive modalities of quantum information processing in the noisy, intermediate-scale quantum era \cite{Preskill:2018aa}.

Reservoir computing \cite{Maass:2002aa,Jaeger:2004aa,Verstraeten:2007aa} is a paradigm in which control of the system is eschewed almost entirely. Instead, information is encoded in signals that stimulate a ``reservoir'', whose inherent nonlinear dynamics and memory serve as computational resources, with only linear post-processing of the reservoir's output required to achieve task-specific performance. While initially implemented in software as computer programs (e.g.,~echo-state networks \cite{Jaeger:2004aa}), reservoirs can also be implemented directly in hardware using complex, nonlinear physical systems whose dynamics need not even be fully understood. Hardware reservoirs built from classical systems have demonstrated high performance at computational tasks \cite{Tanaka:2019aa}, and quantum systems have recently been identified as an attractive choice for a hardware reservoir due to their large state space and complex internal dynamics. Accordingly, the field of quantum reservoir computing (QRC) has received much recent theoretical \cite{Fujii:2017aa,Ghosh:2019aa,Ghosh:2019ab,Ghosh:2020aa,Kutvonen:2020aa,Schuld:2019aa,Chen:2019aa,Nakajima:2019aa,Govia2020}, and experimental \cite{Negoro:2018aa,Chen:2020,Dasgupta:2020} interest.

In this letter, we study the impact of expanding state space on the computational advantage of QRC by studying through simulation the dependence of task performance on the reservoir's Hilbert space dimension ($d$). Compared to the analogous classical reservoir computing (CRC) model, we demonstrate a robust performance improvement that increases with $d$ for QRC at a signal processing task and at the standard memory capacity benchmark for reservoir computing \cite{Jaeger:aa}. When state-space dimension is increased by networking distinct quantum systems, the impact of a larger $d$ can be be confounded by network topology and connectivity strength. To avoid this ambiguity, we consider reservoirs that consist of a single physical element (with a fixed input/output), which allows us to highlight the impact of increasing $d$. By demonstrating flexible, high-performance QRC based on only a single quantum element and clarifying the criteria for successful operation, we cement a paradigm in quantum information processing whose utility will only increase as system complexity matures.

\begin{figure}[!t]
    \centering
    \includegraphics[scale=0.65]{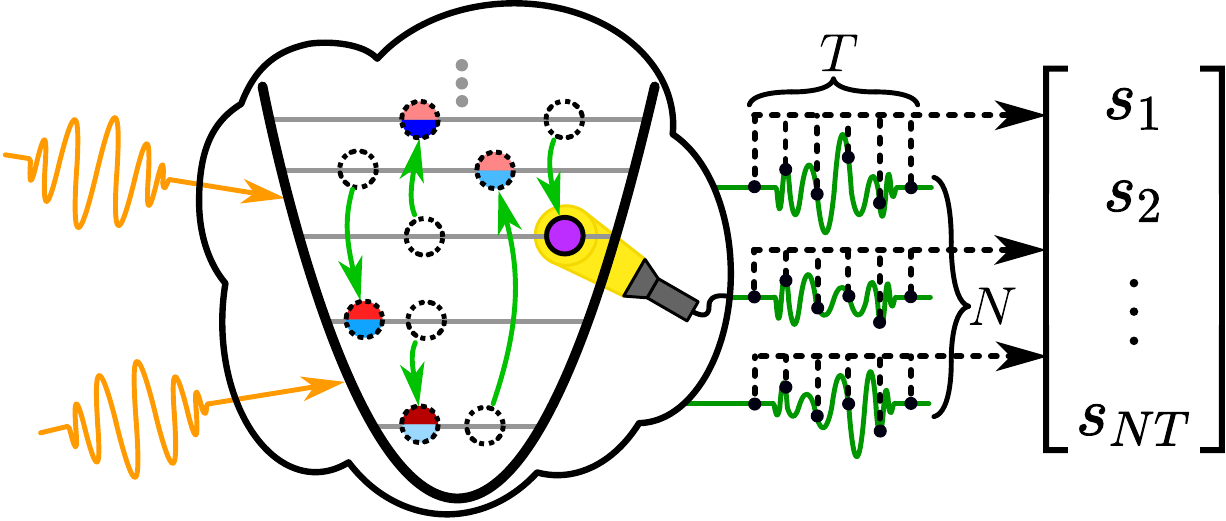}
    \caption{Overview of quantum reservoir computing. Computational nodes are formed by the free variables (represented by red and blue circles) describing a quantum state in Hilbert space. A subset of nodes are stimulated with signals (orange) encoding the input data. In response to this stimulus, the quantum state of the system evolves, and the computational nodes follow an evolution trajectory (green arrows) determined by the reservoir's natural dynamics and the time-dependent input signal. A subset of $N$ nodes are measured (expectation values) at a constant rate in time for $T$ samples. This raw output data is used to perform the desired computation via a final linear post-processing step.}
    \label{fig_setup}
\end{figure}

\emph{Quantum Reservoir Computing---}A general framework for reservoir computing is presented in Fig.~\ref{fig_setup}. The internal dynamics and connectivity of the reservoir are not observed or manipulated, but a subset of nodes that are visible to external interactions form the input/output nodes. In QRC, nodes are the free variables in the density matrix describing the quantum state of the reservoir. More precisely, the state of each node is specified by the expectation value of an element of a complete basis for the operator space on the system's Hilbert space \cite{Fujii:2017aa}.

Reservoir computing is inherently time-dependent, and the raw output vector $\vec{s} \in \mathbb{R}^{NT}$ is a time-series sampling of the output of the $N$ output nodes at $T$ time points in response to an input signal $u(t)$. The sole distinction between classical and quantum reservoir computing lies in dynamical evolution defining the map from input, $u(t)$, to output, $s(t)$
\begin{align}
    s(t) = \mathcal{G}[u(t)].
\end{align}
For CRC, this dynamical evolution is governed by classical physics, while for QRC it is quantum mechanical in nature. In both cases the map $\mathcal{G}$ is fixed, and is not modified during the training procedure.

The desired output vector $\vec{y} \in \mathbb{R}^{L}$ is given by $\vec{y} = \mathbf{W}\vec{s}$, where $\mathbf{W} \in \mathbb{R}^{L \times NT}$ is the output weight matrix for a task with $L$ output classes. Training the reservoir consists of determining $\mathbf{W}$, which is computed as
\begin{equation}
    \mathbf{W} = \mathbf{Y}\mathbf{S}^\intercal \left(\mathbf{S}\mathbf{S}^\intercal + \gamma \mathbb{I}\right)^{-1},
\end{equation}
where $\mathbf{S} \in \mathbb{R}^{NT \times M}$ contains the sample vectors for $M$ training instances, $\mathbf{Y} \in \mathbb{R}^{L \times M}$ contains the known label values, and $\gamma$ is a ridge-regression parameter used to prevent overfitting.

We consider a single $d$-dimensional quantum system (qudit) reservoir, with $d^2-1$ real free variables, with evolution given by the Hamiltonian (we set $\hbar=1$)
\begin{equation}
    \hat{H} = \Omega\hat{a}^\dagger \hat{a} +  K\hat{a}^\dagger \hat{a}\hat{a}^\dagger \hat{a} + u(t)(\hat{a} + \hat{a}^\dagger), \label{eqn:Ham}
\end{equation}
where $\hat{a} = \sum_{n=1}^{d} \sqrt{n}\ketbra{n-1}{n}$ is the qudit lowering operator, $\Omega$ the zero-excitation resonance frequency, and $K$ the nonlinearity. In the limit $d\rightarrow \infty$, the system becomes a single Kerr nonlinear oscillator, but in practice, for large enough $d$ the reservoir effectively models this formally infinite-dimensional system \cite{Govia2020}.

We simulate system evolution using the Lindblad master equation
\begin{equation}
    \dot{\rho} = -i\left[\hat{H},\rho\right] + \kappa\mathcal{D}[\hat{a}]\rho,
\end{equation}
where $\mathcal{D}[x]\rho = x\rho x^\dagger - \{x^\dagger x, \rho\}/2$ is the usual dissipator and $\kappa$ is the rate of excitation decay. Following standard input-output theory \cite{QuantumNoise}, we imagine that the system decays (at least partially) into an output port that allows us to continuously monitor the observable $\hat{X} = \left(\hat{a} + \hat{a}^\dagger\right)/\sqrt{2}$. This gives the continuous raw output signal $s(t) = \Tr[\rho(t)\hat{X}]$, from which we use $T$ sample points to define the sample vector $\vec{s}$ (i.e.~$N=1$).

Throughout this work, we compare performance to the classical analogue of our $d\rightarrow\infty$ quantum model, a Duffing oscillator with an equation of motion
\begin{equation}
    \dot{a} = -i(\Omega+ K)a -i2Ka^2 a^\ast - \frac{\kappa a}{2} - i u(t),
\end{equation}
where $a = (X + i P)/\sqrt{2}$. $X$ and $P$ are the quadratures that constitute the free variables of the reservoir, and we have that $s(t) = X(t)$ as in the quantum model.

\emph{Multivariate Signal Processing---}For our first task, we couple into the reservoir a signal of the form $u(t) = \alpha \sin(\omega t + \phi) + \beta$ and train the output to simultaneously estimate the amplitude $\alpha$ and phase $\phi$ of the signal for unknown constant values of $\omega$ and $\beta$. This is a prototypical signal processing task that demands sufficiently complex internal reservoir dynamics to effectively invert a nonlinear encoding of the phase, and forms a basic building-block for more complicated tasks \cite{Wei00,Serpedin01}. Throughout this task, we will set $\Omega=0$ and measure all frequencies in units of $\kappa$, which we hold fixed.

\begin{figure*}[!t]
    \centering
    \includegraphics[scale=0.8]{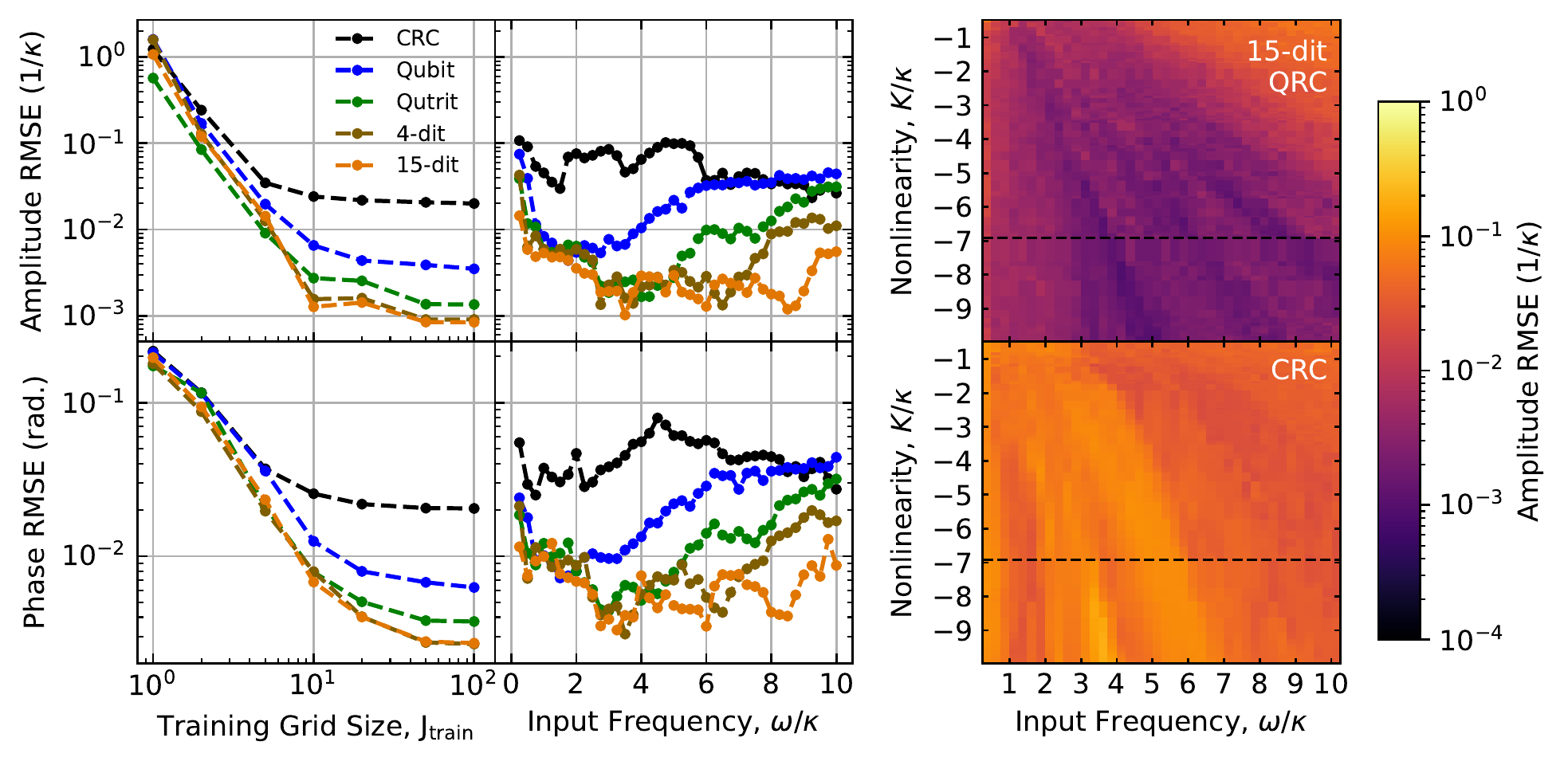}
    \caption{{\bf (Left column)} Dependence of reservoir performance on training set grid-size $J_{\rm train}$ (total training set size $J_{\rm train}^2$) for quantum reservoir computing (QRC) at multiple Hilbert space dimensions, and for classical reservoir computing (CRC). Each data point represents the lowest error for each system when swept over ridge-regression $\gamma$, nonlinearity $K$, and input frequency $\omega$, with a constant static bias $\beta = 10\kappa$. {\bf(Center column)} RMSE as a function of $\omega$ at constant $K$ (shown in the dashed line in the right column), $J_{\rm train} = 100$, and $\beta = 10\kappa$. {\bf (Right column)} Amplitude RMSE as a function of $\omega$ and $K$ for 15-level QRC and for CRC.}
    \label{fig_training}
\end{figure*}

For this task, the raw output vector consists of $T$ samples of the reservoir output over the full duration of each input signal. We train on a $J_{\rm train} \times J_{\rm train}$ grid of uniformly-spaced phases and amplitudes from the intervals $\left[0,\pi/2\right]$ and $\left[\kappa, 10\kappa\right]$, and test performance on $J_{\rm test}$ pairs drawn randomly from uniform distributions over the same intervals. After the total duration of each input signal, we reset the reservoir to its ground state. See \cite{Supplement} for further specifics of the training procedure.

Our metric for performance is the root-mean-squared error (RMSE) on the test set, given by
\begin{equation}
   {\rm RMSE} = \sqrt{\frac{1}{J_{\rm test}}\sum_{j=1}^{J_{\rm test}} (x_j^{\rm est} - x_j^{\rm act})^2},
\end{equation}
where $x_j^{\rm est/act}$ are the estimated and actual phase or amplitude for the $j$'th element of the test set. We evaluate performance for reservoirs with 100 unique values of $K$, randomly chosen from a uniform distribution over the interval $\left[-0.1\kappa,-10\kappa\right]$, and 40 values of signal frequency $\omega$ evenly spaced between $0.25\kappa$ and $10\kappa$. Here, we report the best performance (lowest RMSE) result across these reservoir realizations for each system, with a static bias of $\beta = 10\kappa$ unless otherwise stated (see \cite{Supplement} for complete results).

We start by investigating reservoir performance as a function of training set size (Fig.~\ref{fig_training}) for 2-, 3-, 4-, and 15-level quantum reservoirs, and compare to the classical reservoir. As shown in the Supplementary Information \cite{Supplement}, the higher Fock states in the 15-level system have vanishingly small population, making it a sufficient approximate model of the formally infinite-dimensional quantum harmonic oscillator.

As the left column of Fig.~\ref{fig_training} shows, even the smallest quantum reservoir displays a robust performance advantage over the classical reservoir. All models experience a performance plateau beyond which increasing training set size does not significantly reduce RMSE, and the limit for all QRC sizes is consistently below that of CRC despite similar scaling for small training sets (below 10 samples). Increasing QRC dimensionality gives a monotonic decrease in RMSE, with minimal improvement beyond $d=4$.

Next, we explore how increasing QRC Hilbert space dimension $d$ can improve the high-performance bandwidth. As a function of input frequency $\omega$, and for $J_{\rm train} = 100$, we plot the lowest RMSE from our reservoir realizations (i.e.~best case over $K$) in the middle column of Fig.~\ref{fig_training}. QRC displays an increasing bandwidth as $d$ is increased, and all QRC models have consistently improved performance compared to the CRC model. Further, we find that bandwidth has a strong dependence on reservoir nonlinearity (Fig.~\ref{fig_training} right column). As nonlinearity is increased for the 15-dit quantum reservoir, the high-performance bandwidth grows with increasing $\abs{K}$. In comparison, the opposite is seen for the CRC model, with a widened band of poor performance as $\abs{K}$ increases.

These results demonstrate that even the smallest quantum reservoirs can more effectively exploit nonlinearity as a computational resource compared to a classical reservoir. Further, we have conclusively demonstrated a monotonic increase in QRC performance with Hilbert space dimension, which speaks strongly to the computational utility of a quantum element with complex dynamics that have a high dimensionality to explore.

\begin{figure*}[t]
    \centering
    \includegraphics[width=2.0\columnwidth]{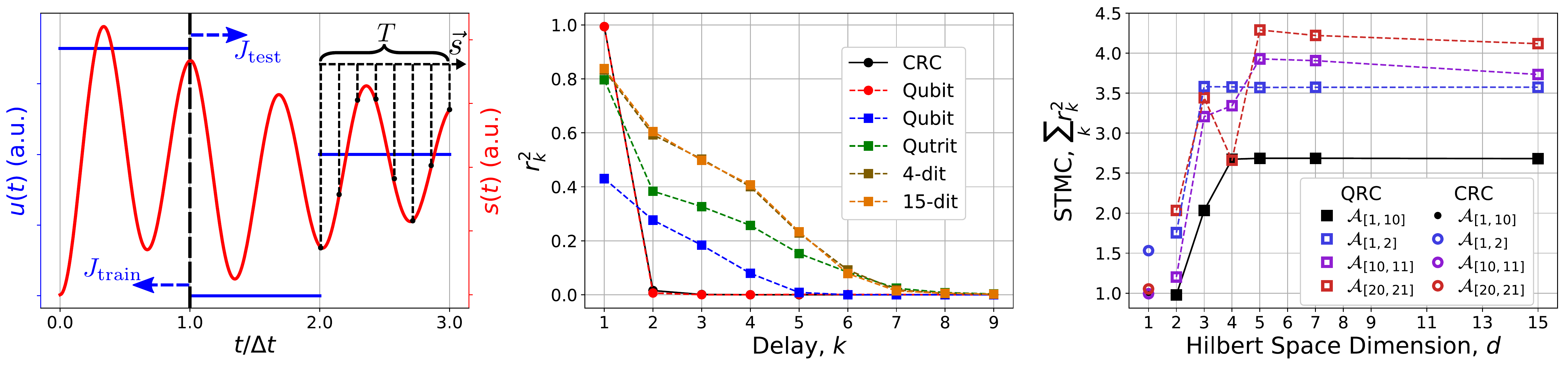}
    \caption{{\bf(Left panel)} Piecewise-constant input, $u(t)$, of the short-term memory capacity task for an example sequence of length $J=3$ ($J_{\rm train} = 1$, $J_{\rm test} = 2$). Also shown is an example output, $s(t)$, and the sampling procedure to generate $\vec{s}_t$, with all $T$ samples taken during the interval $[t,t+\Delta t]$. {\bf(Center panel)} $k$-delay coefficient of determination for QRC models with increasing Hilbert space dimension $d$, and for CRC, with an input amplitude interval $\mathcal{A} = [\kappa,10\kappa]$. For each reservoir model, the curve shown is for the reservoir frequency $\Omega$, hold time $\Delta t$, and measurement sample number $T$ that give the largest STMC value, except for the qubit, where curves for both its best-case $\Delta t$ (red circles) and a $\Delta t$ that matches the other QRC models (blue squares) are shown. {\bf(Right panel)} Short-term memory capacity (STMC) of the QRC models as a function of $d$ for various $\mathcal{A}$. STMC for the CRC models is shown at $d=1$. All results shown are for $K = -50\kappa$, with $J_{\rm train} = 3000$, and $J_{\rm test} = 4000$.}
    \label{fig:MC}
\end{figure*}

\emph{Memory Capacity---}We next study the performance of qudit QRC at the short-term memory capacity (STMC) benchmark to elucidate the impact of Hilbert space dimension on a standard metric of reservoir performance \cite{Jaeger:aa}. Recently applied to QRC qubit models \cite{Kutvonen:2020aa}, this task has not yet been used to study the broad family of multi-level unipartite systems represented by qudit-QRC. STMC tests the reservoir's ability to recall past inputs, therefore serving as a metric for its capacity to store information. This task considers an input $u(t)$ that has a piecewise-constant structure (see left panel of Fig.~\ref{fig:MC})
\begin{align}
    u(t) = \sum_{j=0}^{J-1} \alpha_j\left[\Theta\left(t-j\Delta t\right) - \Theta\left(t-(j+1)\Delta t\right)\right],
\end{align}
where $\Theta(x)$ is the Heaviside step function ($\Theta(0) = 1$) and $\Delta t$ is a fixed hold time. The input samples $\alpha_j$ are uniformly-distributed random numbers drawn from a fixed interval $\mathcal{A}$, and the sample vector $\vec{s}_t$ is constructed from $T$ measurement samples between $s(t)$ and $s(t+\Delta t)$. Note that in contrast to parameter estimation where we sample the reservoir while it evolves under the signal of interest, in the STMC task we gather $T$ output samples from an interval of duration $\Delta t$ that occurs $k\Delta t$ hold times later than when the signal entered the reservoir.

The STMC task determines how well $\vec{s}_t$ can be used to recall the $k$-delayed input $u(t-k\Delta t)$. This is measured by the $k$-delay coefficient of determination
\begin{align}
    r^2_k = \frac{{\rm cov}^2\left[u(t-k\Delta t),\mathbf{W}^k\vec s_t\right]}{\sigma^2\left[u(t)\right]\sigma^2\left[\mathbf{W}^k\vec{s}_t\right]},
\end{align}
which quantifies how much of the variance in the $k$-delayed input can be explained by the variance in the processed output $\mathbf{W}^k\vec s_t$. The output weights $\mathbf{W}^k$ are calculated for each delay time from a continuous input of total time $J\Delta t$, i.e.~$J$ random input values, with the first $J_{\rm train}$ used as the training set, and the remainder as the test set (see \cite{Supplement} for further details).

The short-term memory capacity of the reservoir is then defined as STMC $= \sum_{k=1}^\infty r^2_k$, with larger values indicating improved performance. Assuming i.i.d.~input, the STMC is bounded above by the number of computational nodes in the reservoir \cite{Jaeger:aa}. We calculate the STMC of the CRC and QRC models for a variety of hold times $\Delta t$, frequencies $\Omega$, nonlinearities $K$, training set sizes $J_{\rm train}$, and amplitude intervals $\mathcal{A}$.

We observe that the STMC improves (but eventually saturates) with $J_{\rm train}$, has minimal dependence on $\Omega$, and generally increases with $|K|$ \footnote{In Fig.~\ref{fig:MC} we use a larger nonlinearity than in the previous task to emphasize the separation between CRC and QRC}. It exhibits a strong dependence on hold time, with a clear divide between CRC and QRC. For QRC with $d\geq 3$, the best STMC was found for a hold time of $\Delta t = 0.5/\kappa$, while $\Delta t = 5/\kappa$ gave the best STMC for CRC and QRC with $d = 2$. As pointed out in Ref.~\cite{Govia2020}, QRC with $d=2$ is equivalent to a classical reservoir model with additional nonlinear output processing. The order-of-magnitude separation in hold time for best STMC is evidence that there is a distinct operational difference between ``true'' quantum reservoirs (QRC $d\geq3$) and CRC. In the remainder of this section we show STMC results for best-case timing, with further details in the supplementary material \cite{Supplement}.

The center panel of Fig.~\ref{fig:MC} shows $r^2_k$ as a function of delay, with $\mathcal{A} = [1,10]$ chosen to match the range used in the signal processing task. The decay of $r^2_k$ strikingly demonstrates the effect of the different optimal $\Delta t$, with longer-lived $r^2_k$ for the $d\geq3$ QRC models. The hold time determines the shape of this decay, as seen by comparing qubit $r^2_k$ at its best-case $\Delta t = 5/\kappa$ (red circles), to that for $\Delta t = 0.5/\kappa$ (blue squares). For the longer $\Delta t$ the decay shape is indistinguishable from that of CRC, while for the shorter $\Delta t$ it is the same as that of $d\geq 3$ QRC. While the decay shape is determined by hold time, we emphasize that the magnitude of $r^2_k$ shows a clear performance improvement for QRC with increasing $d$.

As the right panel of Fig.~\ref{fig:MC} shows for a variety of $\mathcal{A}$ (chosen to be representative of the effects of $\mathcal{A}$ on STMC \cite{Supplement}), the STMC for QRC grows with Hilbert space dimension, saturating beyond $d=4$. The saturation values are all well above the largest STMC found for CRC, which further demonstrates the robust performance advantage and improvement with $d$ of QRC.

Nevertheless, the STMC we have shown is far from the upper bound set by reservoir dimension. STMC is limited by: i) the number of independent reservoir variables (controlled by the input scheme and internal dynamics) and ii) access to independent observables. Thus, our results are likely limited by the insufficiently complex evolution of Eq.~\eqref{eqn:Ham}, which is unable to drive transitions throughout the entirety of Hilbert space, as well as by the single-variable output scheme. The QRC models and input/output scheme of this letter were chosen with ease of experimental implementation in mind; the future development of QRC will have to carefully balance ease of implementation with computational power, with one straightforward resolution being the engineering of more complex internal dynamics. It is worth noting that similar concerns affect CRC, as we have seen in this work for the STMC of the CRC model.

\emph{Conclusion---}In this work, we have studied quantum reservoir computing using single-qudit reservoirs representative of current-era quantum hardware. Compared to the $d\rightarrow \infty$ analogous CRC model of a Duffing nonlinear oscillator, we find a robust improvement in performance for all QRC models at both a signal processing task and the short-term memory capacity benchmark. Moreover, we demonstrate a monotonic improvement in QRC performance with Hilbert space dimension, providing strong evidence for the notion that Hilbert space acts as computational space in a quantum reservoir.

Our results further show that not all of Hilbert space is straightforwardly accessible as computational space, and careful consideration of the input/output scheme and internal dynamics is necessary to take advantage of the full power of quantum reservoir computing. The future success of QRC will depend on accessing an exponentially-expanding potential computational space, which may require clever architectural design of  interactions with the reservoir, as well as the reservoir itself.

The QRC models we consider are readily implementable in contemporary physical platforms. For example, qudits can be directly realized in trapped ions \cite{Low2020}, molecules \cite{Sawant:2020aa}, or superconducting circuits \cite{Krantz:2019aa}. Similarly, strongly nonlinear oscillators have been demonstrated in superconducting \cite{Yamaji2020}, optomechanical \cite{Sankey:2010aa}, and photonic \cite{Strekalov_2016} systems. Given the generality of our models, they can find immediate application in almost any contemporary quantum information processing platform.

\acknowledgements
\emph{Acknowledgments---}This material is based upon work supported by the U.S. Army Research Office under Contract No: W911NF-19-C-0092. Any opinions, findings and conclusions or recommendations expressed in this material are those of the authors and do not necessarily reflect the views of the U.S. Army Research Office.

\bibliography{QRC_Qudit_combined.bib}

\clearpage
\clearpage

\setcounter{figure}{0}
\renewcommand{\thefigure}{S\arabic{figure}}
\renewcommand{\theHfigure}{Supplement.\thefigure}

\onecolumngrid
\section{Supplementary Information for ``Neuromorphic computing with a single qudit''}

\twocolumngrid
\section{Training/Testing and Reservoir Realizations}

In our simulations, we have explored a wide range of reservoir realizations, with each QRC realization a qudit with different internal parameters ($\Omega, K$), as well as input parameters ($\omega, \beta$). Training is performed for each realization separately by

\begin{enumerate}
    \item Generating the elements of the training set for the specific task (as described below). We use the same training and testing sets for all realizations.
    \item Simulating the evolution of the reservoir under the corresponding input signal for each training and test sample.
    \item Calculating the value of an observable for $T$ uniformly-spaced time points (choice of $T$ and total sampling time is described later in the supplement) for each training sample.
    \item Constructing $\mathbf{S}$, whose columns are the time-series traces for each training sample.
    \item Computing $\mathbf{W}$ as given by Eq.~(1) of the main text.
\end{enumerate}

When training, we sweep the ridge regression parameter $\gamma = \pm 10^p$, where $p$ ranges from -12 to 0 in integer steps, to find the value that gives the best performance on the test set. Ridge regression is used to prevent over-fitting of the training data, and negative ridge regression can be beneficial when the output contains many degrees of freedom uncorrelated with the task objective \cite{Kobak2018}. Note that for the short-term memory capacity benchmark we only use positive $\gamma$ to be more aligned with standard conventions.

For each realization, testing consists of calculating the relevant test metric for the test set: RMSE (Eq.~(5) of the main text) for multivariate signal processing, or $k$-delay coefficient of determination (Eq.~(7) of the main text) for the short-term memory capacity.

\subsection{Multivariate Signal Processing}

For this task, the training set size is described by a number $J_{\rm train}$, such that $J_{\rm train}$ uniformly-spaced amplitudes and $J_{\rm train}$ uniformly-spaced phases are used to generate a grid of input parameters, for a total training set of size $M = J_{\rm train}^2$ unique elements. During each task we evaluate performance on $J_{\rm test} = 3000$ test samples, each of which is a phase-amplitude pair.

We evaluate performance for reservoirs with 100 unique values of $K$, randomly chosen from a uniform distribution over the interval $\left[-0.1\kappa,-10\kappa\right]$, and 40 values of signal frequency $\omega$ evenly spaced between $0.25\kappa$ and $10\kappa$. This yields an amplitude RMSE with units of energy (which we express as a fraction of $\kappa$) and phase RMSE with units of radians.

\subsection{Short-Term Memory Capacity Benchmark}

As described in the main text, we generate a continuous, piecewise-constant signal of varying amplitude and of length $J \Delta t$. This is spliced to form a set of $J_{\rm train}$ training samples consisting of the first $J_{\rm train}$ amplitudes, i.e.~the initial $J_{\rm train}\Delta t$ duration of the input pulse. We report performance on the remaining $J_{\rm test} = J - J_{\rm train}$ input samples, which form the test set. In our simulations, we report results for $J_{\rm train} = 3000$ and $J_{\rm test} = 4000$ in the main text.

We evaluate the STMC benchmark for reservoir realizations with $K$ taking a value from $ [-10\kappa,~-20\kappa,~-50\kappa]$, and $\Omega$ a value from $[0,~\kappa,~2.5\kappa]$. As discussed in the main text, we find that performance improves with $K$ for all reservoirs, but does not strongly depend on $\Omega$. We use $\Omega = 0$ for the classical reservoir, and $\Omega = \kappa$ for the quantum reservoir results shown in the main text, as these values gave the highest STMC.

\section{Sampling Optimality}

Reservoir performance depends strongly on the output duration, and the number of measurement samples. We now discuss our strategy to find values of these parameters that result in high performance for our chosen tasks.

\subsection{Multivariate Signal Processing}

\begin{figure}[t]
    \centering
    \includegraphics[scale=0.7]{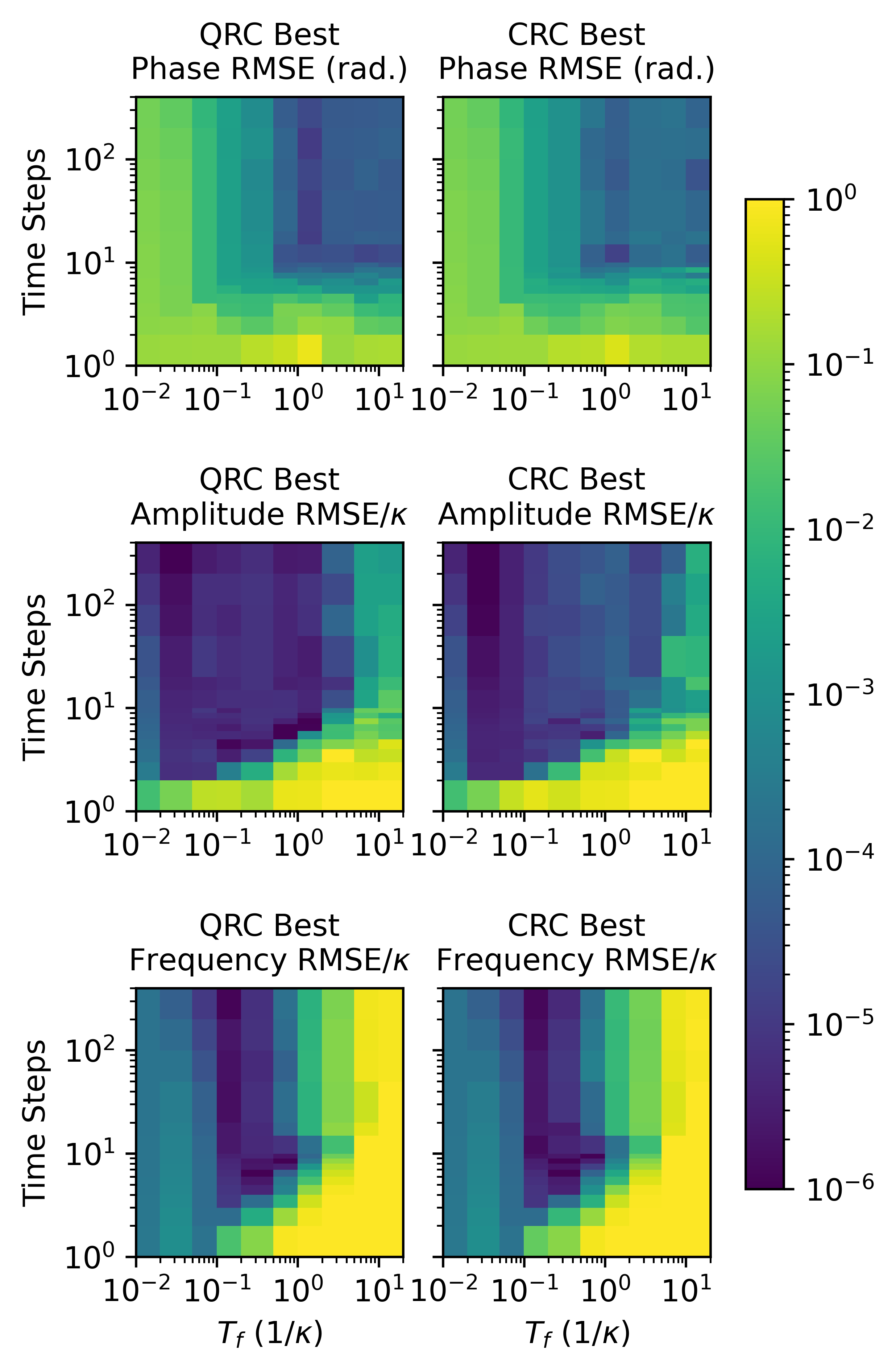}
    \caption{Reservoir performance as a function of sampling parameters for the 15-dit quantum reservoir (QRC) and the classical reservoir (CRC). Best-case performance (i.e.~lowest RMSE) across reservoir realizations is shown.}
    \label{fig_timing}
\end{figure}

Before attempting to simultaneously learn multiple parameters of a signal, we test the ability of the 15-level qudit and classical reservoir to learn either the amplitude, phase, or frequency while holding the other two parameters fixed. We then sweep the fixed parameters as well as $K$ and note the best-case performance. We repeat this as we vary the length of time over which we sample the reservoir output as well as the number of samples captured in that time, and report the results in Fig.~\ref{fig_timing}.

We note that the quantum and classical reservoirs exhibit similar temporal structure, which is unique for each task, yet have distinct performance optima. Striking a balance between the optimal timing parameters for separate phase and amplitude estimation, we choose a total sample time $T_f = 2/\kappa$ and $T = 51$ sample points for all quantum reservoirs and $T_f = 0.5/\kappa$ and $T = 21$ for the classical reservoirs.

\begin{figure*}[t]
    \centering
    \includegraphics[scale=0.7]{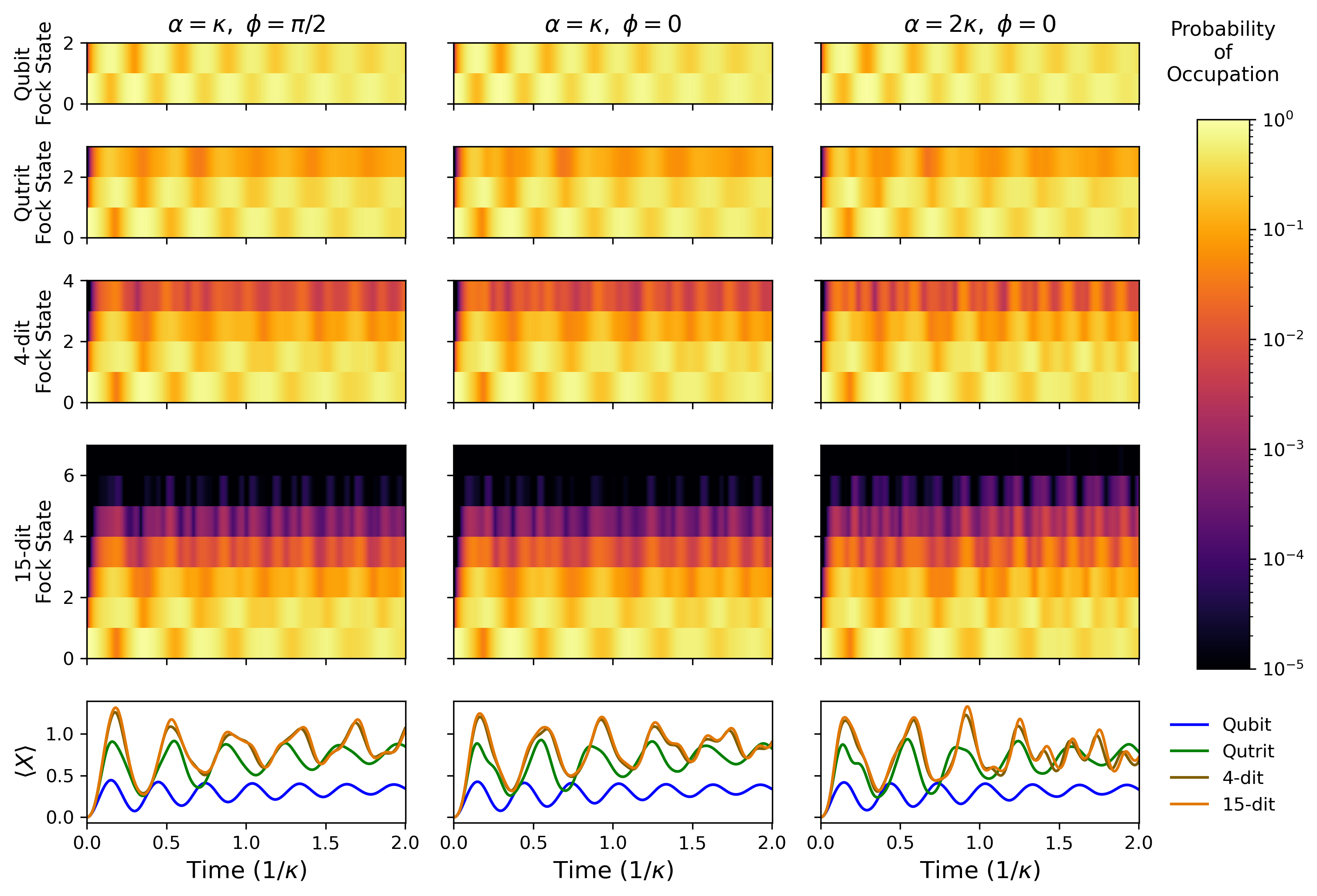}
    \caption{Fock state occupation probabilities and $X$-quadrature expectation values for QRC systems with $K=-7\kappa$ during time evolution under an input signal with $\omega=6\kappa$ and $\beta=10\kappa$. Each column corresponds to different values of $\alpha$ and $\phi$, and the effects of changing these signal parameters are reflected in the evolution of the system.}
    \label{fig_state}
\end{figure*}

Example quantum state trajectories for three different pairs of signal amplitudes and phases are presented in Fig. \ref{fig_state}. The expectation value of the $X$-quadrature, displayed in the bottom row, exhibits more complex behavior for models with higher dimensionality due to the occupation of higher Fock state levels combined with the system's nonlinear evolution. Hence, changing the amplitude and phase of the input signal leads to more drastic changes in the time evolution of the system. As highlighted in the main text, the occupation of Fock state $m$ in the 15-dit model is already below $10^{-4}$ by $m=6$. This indicates that this model is indeed a good approximation to the infinite-dimensional nonlinear oscillator, as there is no quantitative impact of the finite Fock state cutoff on simulation dynamics.

\subsection{Short-Term Memory Capacity Benchmark}

For this task, the input and output duration are locked together, and are set by the input sample duration $\Delta t$. As discussed in the main text, there is qualitatively and quantitatively different performance for the classical and quantum reservoirs depending on the value of $\Delta t$. The performance dependence on the number of measured sampling points, $T$, in a given interval $\Delta t$ is considerably weaker. For this task, $\Delta t$ and $T$ were treated equivalently to the other system parameters, $\Omega, K, \omega$, and course-grained searches were done to find high-performing reservoir realizations. The best-case $\Delta t$ for the classical and quantum reservoirs reported in the main text were found by evaluating performance across a range of $\Delta t$ from $\Delta t = 0.1/\kappa$ to $\Delta t = 6/\kappa$. For all reservoirs, the best performance was found for $T = 100$ sampling points, with $T = 50,~200$ also tested.

\section{Numerical Simulations}

We simulate quantum and classical reservoir evolution by numerically solving Eqs.~(3) and (4) of the main text in the Julia Language \cite{Bezanson:2017aa}. For the simulations of the multivariate signal processing benchmark we use custom GPU-accelerated simulation software that parallelizes parameter sweeps across an NVIDIA GTX 1080 Ti GPU. For each simulation, this software employs the Tsitouras 5/4 Runge-Kutta method included within {\tt DifferentialEquations.jl} \cite{DifferentialEquations.jl-2017} to solve the relevant differential equation. These computations are carried out with 64-bit floating point arithmetic, as we observed that using 32-bit arithmetic led to an appreciable accumulation of error at the end of the total simulation time (as determined by comparisons to CPU simulations that used 64-bit arithmetic). The short-term memory capacity simulations were carried out on a CPU using {\tt MESolve.jl} \cite{MESolve}, a time-dependent master equation solver written in Julia.

\section{Supplementary Simulation Results}

\subsection{STMC Training and Test Set Size Dependence}

For the short-term memory capacity benchmark, sufficient training is required for the reservoir to achieve high performance, and a sufficiently large training set must be used to ensure that anomalously high performance is not reported due to finite statistics. However, increasing simulation time sets a practical constraint on the total length of an input sequence that can be considered.

\begin{figure}[t]
    \centering
    \includegraphics[width=\columnwidth]{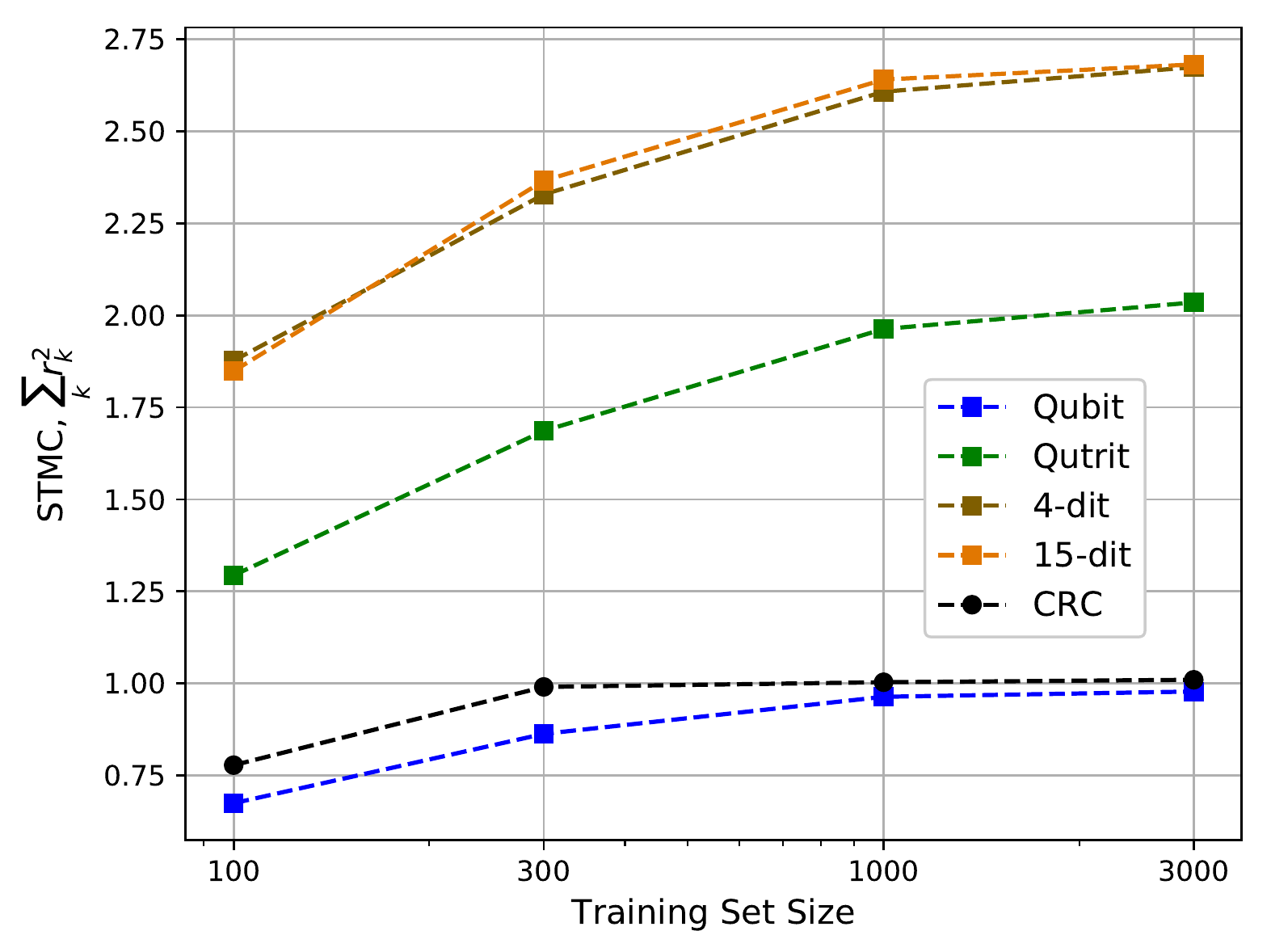}
    \caption{Short-term memory capacity for the CRC and QRC models as a function of training set size. For each data point, the test set size varied as $J_{\rm test} = 7000 - J_{\rm train}$. Reservoir and input parameters are the same as used in the main text.}
    \label{fig:MC_Sup1}
\end{figure}

We explored the increase in STMC as a function of training set size, the results of which are shown in Fig.~\ref{fig:MC_Sup1}. As can be seen, for all the models considered the STMC has saturated (or nearly saturated) by the time $J_{\rm train} = 3000$. As such, we can expect that the results shown in the main text, which use $J_{\rm train} = 3000$, cannot be improved upon significantly by increasing training set size.

\begin{figure}[t]
    \centering
    \includegraphics[width=\columnwidth]{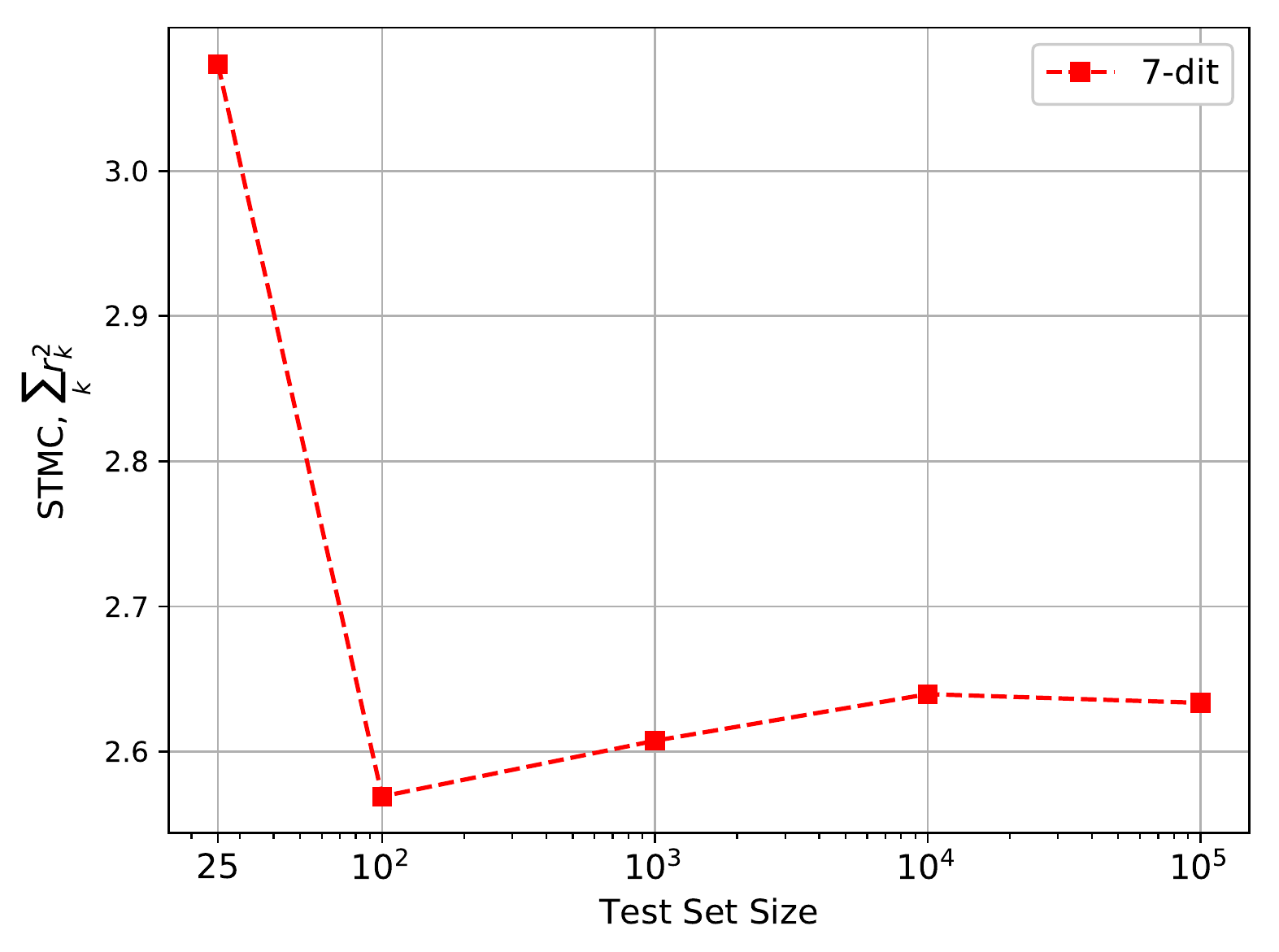}
    \caption{Short-term memory capacity for a 7-dit QRC model as a function of test set size. For each data point, the training set size was held constant at $J_{\rm train} = 3000$. Reservoir and input parameters are the same as used in the main text.}
    \label{fig:MC_Sup2}
\end{figure}

Similarly, we consider varying the test set size, shown in Fig.~\ref{fig:MC_Sup2} for 7-level QRC. To ensure that we are not overestimating performance, we consider test set sizes up to two orders of magnitude larger than the $J_{\rm train} = 4000$ used in the main text. As can be seen in Fig.~\ref{fig:MC_Sup2}, once the test size is beyond order 100, variance in the STMC is small, and can be explained by fluctuations due to finite sample size effects.

\subsection{Multivariate Signal Processing Parameter Sweeps}

\begin{figure*}[t]
    \centering
    \includegraphics[scale=0.6]{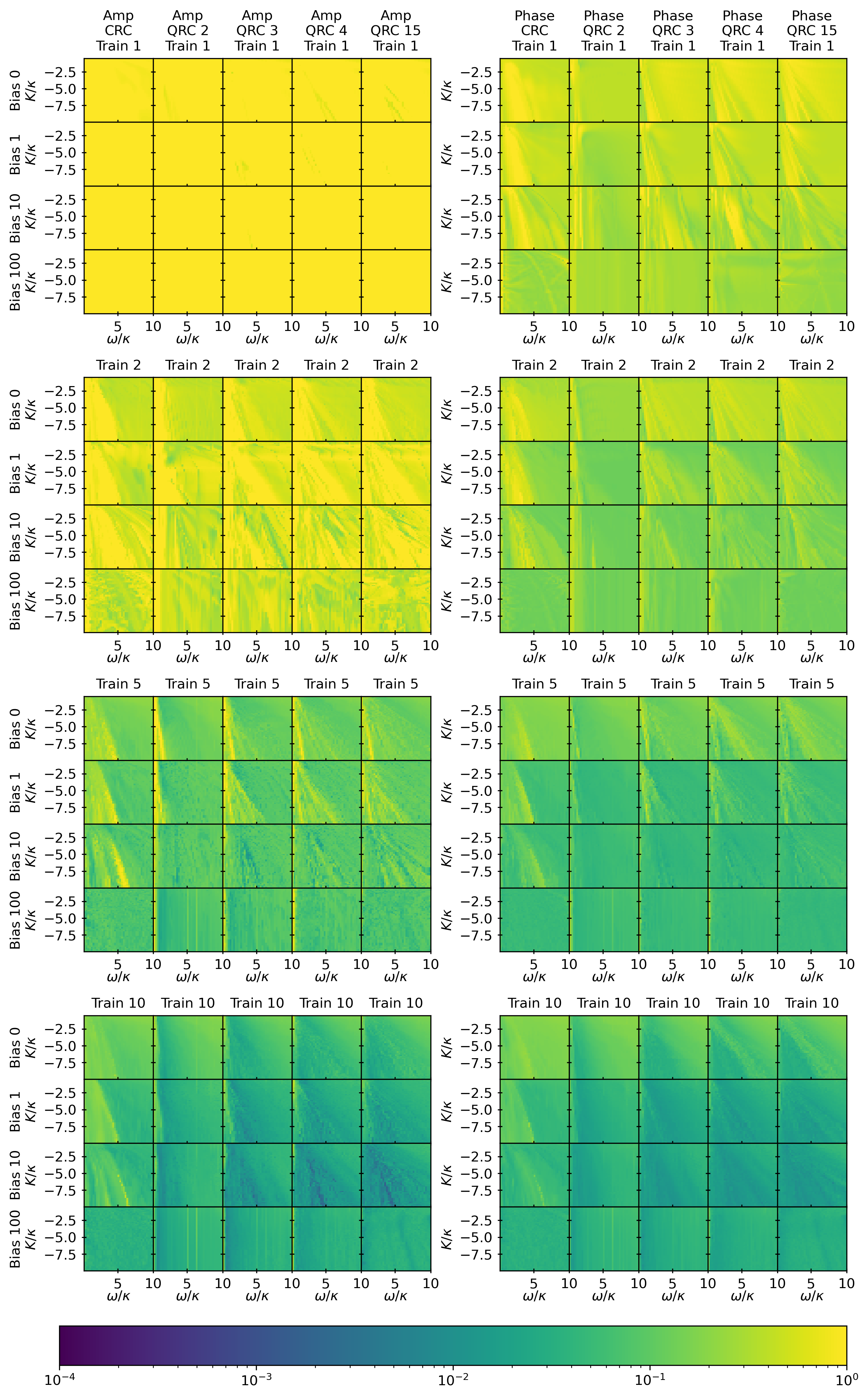}
    \caption{Amplitude and phase estimation RMSE as a function of nonlinearity $K$, input frequency $\omega$, and static bias $\beta$, for the various classical and quantum reservoir models. This figure shows training set sizes of 1, 2, 5, and 10 element grids.}
    \label{fig_all1}
\end{figure*}

\begin{figure*}[t]
    \centering
    \includegraphics[scale=0.6]{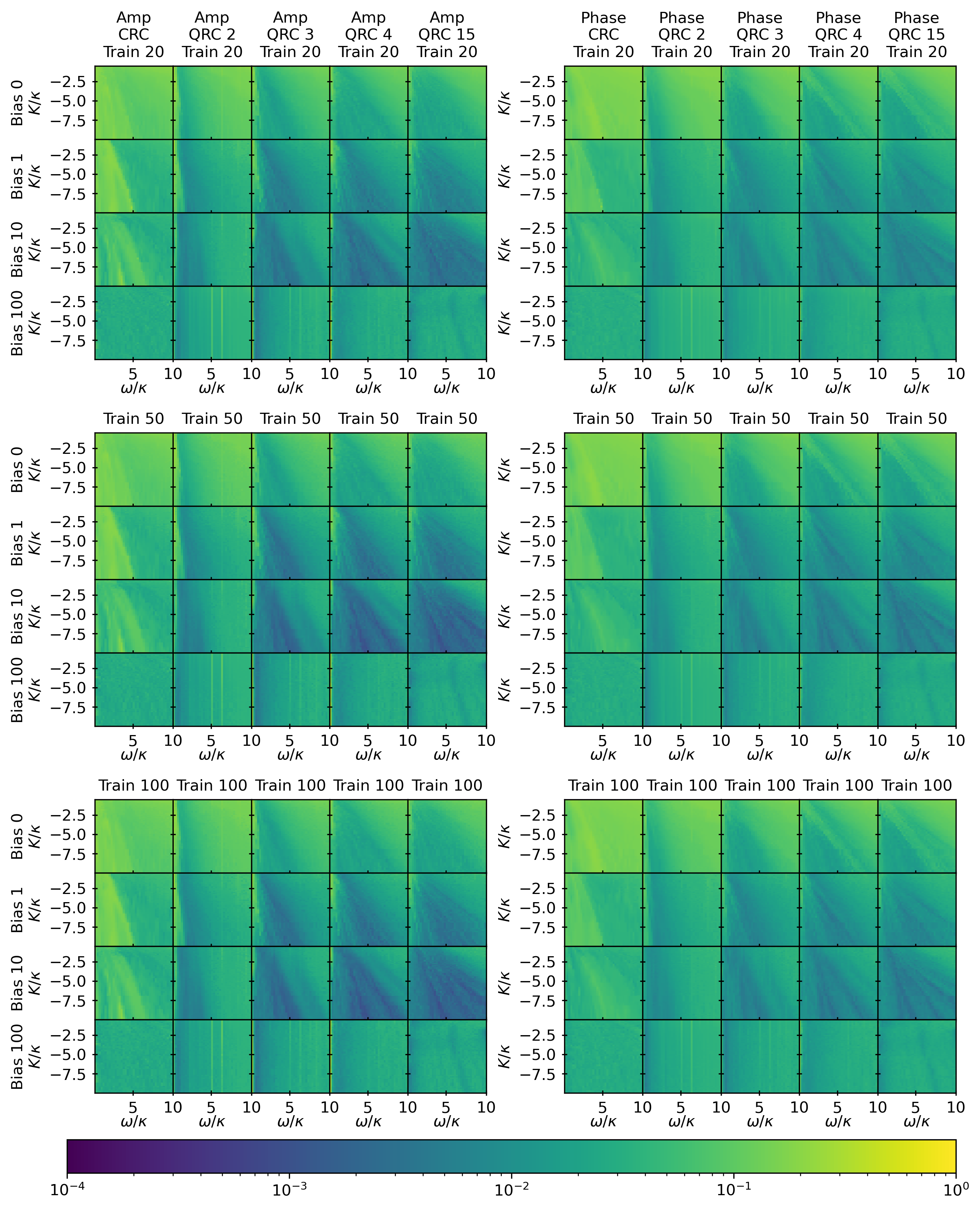}
    \caption{Amplitude and phase estimation RMSE as a function of nonlinearity $K$, input frequency $\omega$, and static bias $\beta$, for the various classical and quantum reservoir models. This figure shows training set sizes of 20, 50, and 100 element grids.}
    \label{fig_all2}
\end{figure*}

The complete set of simulation results for this task, the RMSE as a function of the reservoir parameter sweeps, is presented across Fig. \ref{fig_all1} and Fig. \ref{fig_all2}.

\end{document}